# Standoff Detection of Solid Traces by Single-Beam Nonlinear Raman Spectroscopy Using Shaped Femtosecond Pulses

O. Katz[1], A. Natan[1], S. Rosenwaks[2] and Y. Silberberg[1]

[1]Department of Physics of Complex Systems, Weizmann Institute of Science, Rehovot, 76100 Israel

[2]Department of Physics, Ben Gurion University of the Negev, Beer Sheva 84105, Israel.

E-mail: yaron.silberberg@weizmann.ac.il

*We demonstrate a single-beam, standoff (>10m) coherent anti-Stokes Raman scattering spectroscopy (CARS) of various materials, including trace amounts of explosives and nitrate samples, under ambient light conditions. The multiplex measurement of characteristic molecular vibrations with <20cm$^{-1}$ spectral resolution is carried out using a single broadband (>550cm$^{-1}$) phase-shaped femtosecond laser pulse. We exploit the strong nonresonant background signal for amplification of the weak backscattered resonant CARS signal by using a homodyne detection scheme. This facilitates a simple, highly sensitive single-beam spectroscopic technique, with a potential for hazardous materials standoff detection applications.*

Recently, there has been a growing interest in the challenge of remotely detecting and identifying hazardous materials such as chemical and biological warfare agents and explosives at a standoff distance [1-4]. The Raman vibrational spectrum of molecules provides an excellent fingerprint for species identification, and can be harnessed for this task [2-6]. In multiplex CARS, a band of vibrational levels are excited by a broadband pump and/or stokes beam, and subsequently probed by a narrow frequency probe beam, $\omega_{pr}$ [4-5,7-8]. The vibrational level energies, $\hbar\omega_{vib}$ are resolved by measuring the amount of blue-shift of a scattered Anti-Stokes photon energy from the probe photon energy, $\omega_{AS}=\omega_{pr}+\omega_{vib}$. With the advent of powerful ultrafast lasers, CARS has become the method of choice in nonlinear optical spectroscopy and microscopy [5,9,10]. However, in most cases where short pulses are utilized for CARS spectroscopy, the resonant CARS signal from the vibrational level of interest is obscured by the much stronger nonresonant FWM signal [5-8].

A variety of CARS spectroscopy schemes have been developed in order to suppress the nonresonant background [6,11-13]. These schemes employ various techniques such as tailoring the probe-pulse temporal width and delay [6,11], polarization shaping [12] and quantum coherent control [11,13], exploiting different properties of the nonlinear four-wave-mixing interaction to achieve the goal of background-free measurements [12]. While these novel techniques are constructive for applications such as chemical imaging [13-14], they are not necessarily optimal for long-range standoff probing of scattering samples. Under these conditions, polarization manipulation techniques [2,12] are sensitive to the depolarization caused by the multiple random scattering in the sample, and techniques that effectively suppress the nonresonant signal [6,11,13], might results in a resonant signal which is too weak to be detected.

In this work we employed a single-pulse phase-contrast multiplex CARS technique [7], for long-range standoff probing of minute amounts of scattering samples. In this technique, a broadband, ultrashort pulse supplies both the broadband pump and Stokes photons, and a narrow-band portion of the same pulse is phase-shifted to serve as the probe beam. Furthermore, instead of struggling in reducing the strong nonresonant background, it is exploited for the amplification of the weak resonant signals by interfering with them coherently. The amplification of the resonant CARS signal is carried through a homodyne detection scheme, as is evident from the expression for the measured CARS signal intensity [8]:

$$I(\omega) = |P_R(\omega) + P_{NR}(\omega)|^2 = |P_R(\omega)|^2 + |P_{NR}(\omega)|^2 + 2P_{NR}(\omega)\mathrm{Re}[P_R(\omega)] \qquad (1)$$

where $I(\omega)$ is the measured signal intensity at the frequency $\omega$, $P_R$ and $P_{NR}$ are the resonant and nonresonant CARS signals, respectively. Under typical experimental conditions with ultrashort pulses, the nonresonant signal is considerably stronger than the resonant signal, $|P_{NR}|^2 >> |P_R|^2$, making the latter difficult to detect [6-8]. However, since the nonresonant signal is usually a smooth pure-real function in frequency [7,8], the cross term in Eq. (1), $P_{NR}(\omega)\mathrm{Re}[P_R(\omega)]$, has the shape of the real part of the resonant signal, multiplied by the much stronger nonresonant signal. As a result, this term yields a distinctive amplified feature which is linear in the resonant signal amplitude [7]. The combination of the considerable amplification in resonant signal and the polarization insensitivity within a multiplex CARS scheme, make it attractive for stand-off probing applications [1].

To demonstrate the ability to measure backscattered CARS signals from a standoff distance, this single-pulse multiplex CARS scheme was implemented experimentally using a broadband ultrafast (30fs) laser, phase-controlled by a pulse shaper [15] (Fig 1a). The single broadband (>550cm$^{-1}$) pulse served as both the broadband pump and Stokes pulses, and as the source for a much narrower (<20cm$^{-1}$) probe, which determined the multiplex CARS spectral resolution [7]. The narrowband probe was defined within the pulse spectrum by shifting its spectral phase by $\pi$ in a narrow frequency range, applying the phase-mask shown in Fig 1b. This shape will be referred to as a $\pi$ phase gate. To enable the measurement of the blue-shifted CARS spectrum, the short-wavelength end of the pulse was suppressed by a long-pass filter and a variable knife-edge slit located in the Fourier plane of the pulse-shaper (Fig 1). As a result of the phase-gate, the resonant and nonresonant signal interfere constructively at the high frequency side of the gate, generating a peak in the CARS spectrum, and destructively at the low frequency side, generating a dip in the CARS spectrum (Fig. 2a) [7]. High-resolution standoff spectroscopy was achieved by subtracting the CARS spectra obtained with a transform limited, flat-phase pulse, from that obtained with the π phase-gated pulse. The vibrational Raman spectrum is easily extracted from the spectral location and amplitude of this peak-and-dip feature, using a matched-filter (Fig. 2).

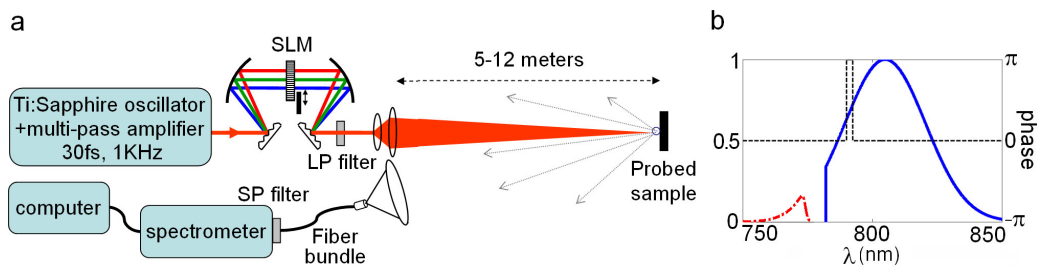

Fig. 1: (a) The experimental setup. The laser source is an amplified Ti:Sapphire laser with a dispersion compensating prism compressor (Femtolasers GmbH). The pulses (0.5mJ, 30fs at 1KHz repetition rate) are phase-shaped in a pulse shaper using an electronically controlled liquid-crystal spatial light modulator (Jenoptik Phase SLM-640). The short wavelength end of the spectrum is suppressed using a long-pass (LP) interference filter and by amplitude shaping using a variable knife-edge slit at the shaper's Fourier plane. The beam is focused on the distant sample through a telescope, and the scattered radiation is collected with a 7.5" diameter lens. The collected light is short-pass filtered and is fiber-coupled to an imaging spectrometer with a liquid Nitrogen cooled CCD (Jobin Yvon Triax 320). (b) Illustration of the spectral amplitude (solid blue) and phase (dashed black) of the phase-gated laser pulse. The π phase-gate selects the narrow probe from the wide bandwidth of the pulse; Also shown is the detected CARS signal spectral region (dash-dot red).

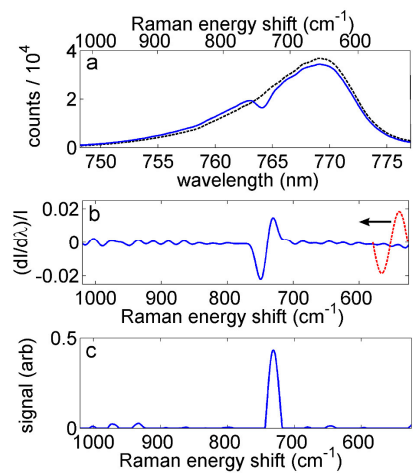

Fig. 2: Extraction of the vibrational level spectrum from the measured CARS signal: (a) Measured CARS signals from bulk PTFE (Teflon) at a standoff distance of 5m, for a transform limited pulse (dashed black) and for a phase-gated shaped pulse (solid blue). The distinctive peak-dip interference feature of a resonant vibrational level is apparent at 764nm; (b) Normalized intensity derivatives difference of the measured CARS spectra (blue), compensate for the changing local-oscillator intensity, and yields a signal which is proportional to the resonant CARS signal (Eq. 1); A convolution of this signal with a pre-determined matched filter (dashed red), extracts the vibrational levels spectrum; (c) The resolved vibrational spectrum reveals the strong 732cm$^{-1}$ line of Teflon. The integration time is 600ms and the phase-gated probe is located at 810nm (not shown).

Experimentally resolved vibrational spectra of minute amounts of solids, liquids and explosives are presented in Fig. 4. All of the spectra were obtained in ambient light conditions with a black absorber screen placed behind the sample. The absorber placed behind the samples ensured the collection of only the weak backscattered signal, avoiding possible reflection of the strong forward CARS signal which is much easier to detect [5]. The measured spectra are in good agreement with the known vibrational spectra of the probed materials [16,17]. The average power on the samples was 90-140mW, yielding peak intensity of ~$3\cdot10^{11}$ Watt/cm$^2$.

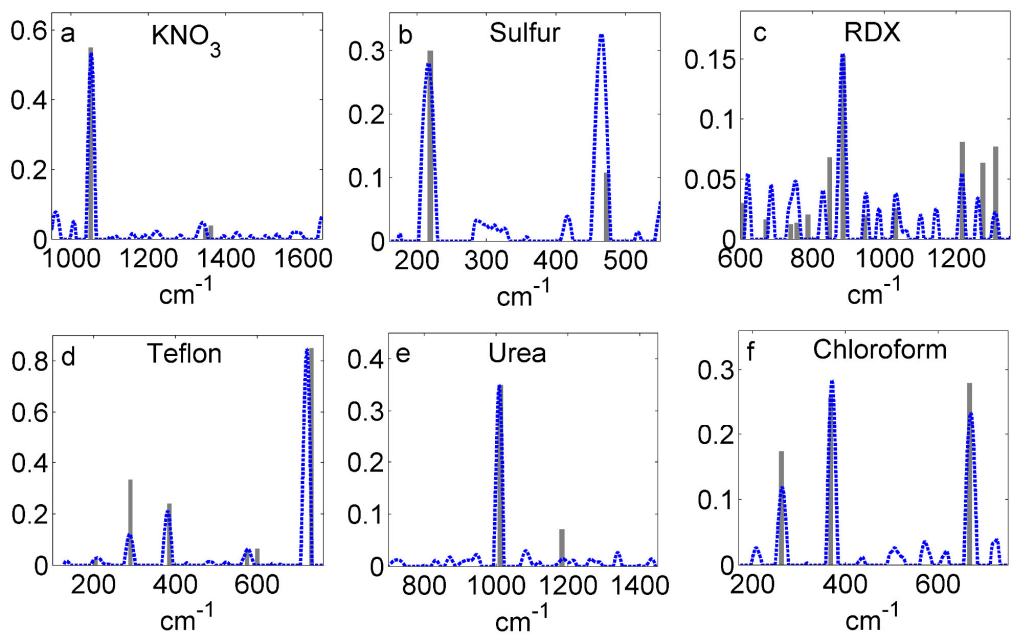

Fig. 3: Resolved femtosecond CARS vibrational spectra of several scattering samples (dashed blue), obtained at standoff distances of 12m (a-c) and 5m (d-f): (a) <1000μg crystallized KNO$_3$; (b) <500μg Sulfur powder; (c) RDX explosive particles with a total mass of <4mg; (d) Bulk PTFE; (e) <4mg of crystallized Urea particles; (f) 1cm long cuvete containing Chloroform and scattering ZnTe particles (200nm diameter); Each spectrum was resolved from a single measurement with an integration time of: (a)-(c) 3 seconds (d) 1 second; (e) 300msec; (f) 350msec. The Raman vibrational lines and their relative strengths are plotted by gray bars for comparison [16,17].

CARS signals are cubically proportional to the intensity, yielding stronger signals for tightly focused beams. In our experiments the minimal beam diameter of ~1-2mm was limited by the beam-point stability of the experimental setup. Actively stabilizing the beam should allow tighter focusing, resulting in a stronger CARS signal and reduced noise from variations between measurements. In our single-beam technique spatial overlap of the pump, probe and Stokes photons is automatically fulfilled. However, assuring temporal overlap of the entire pulse bandwidth at large distances requires dispersion compensation. Dispersion compensation can be achieved by applying an appropriate phase-mask using the pulse shaper [18,19], or as was done in our experiments, by using the integral prism compressor of the laser source.

Further improvements can be obtained: the resolved vibrational spectral coverage can be extended by spectral broadening of the laser source [20]. Shorter pulses with wider bandwidth would also increase peak power and thus the CARS signal. Gated detection (e.g. by the use of an intensified CCD [3]) will increase the signal-to-noise (SNR) considerably. Furthermore, The coherent nature of our technique allows tailoring of the pulse shape for enhanced selectivity when the vibrational spectrum of the suspected contaminants is known [3,6,21]. This can be achieved by using multiple appropriately located phase-gated probes to generate a large coherent spectral feature from the constructive interference of the different vibrational levels [21]. This approach which is not accessible in conventional spectroscopic methods holds the potential for a considerable improvement in the SNR and a reduction in detection integration time.

Femtosecond CARS spectroscopy exhibits higher efficiency at low average powers compared to longer (nsec) pulses used in conventional CARS and Raman techniques [5,9]. This is a merit for the nondestructive probing of sensitive samples such as explosive materials, that may deteriorate under intense illumination [3].

Our experiments demonstrate that coherently controlled femtosecond CARS can be applied to high sensitivity spectroscopy, and in particular for the detection of minute amounts of complex molecules at standoff distances. This technique is attractive for standoff detection of hazardous materials, such as explosives and chemical agents, an issue of obvious practical importance.

**Acknowledgements**

We thank Dan Oron and Nirit Dudovich for fruitful discussions and Keren Kantarovich for the confocal Raman spectra. S.R. is the incumbent of the Helen and Sanford Diller Family Chair in Chemical Physics. Y.S. is the Harry Weinrebe Professor of Laser Physics.